\documentclass[aps,10pt,final,
notitlepage, oneside, nobibnotes, nofootinbib,
superscriptaddress,  centertags, assume]{revtex4}

\usepackage{epsfig}
\usepackage{graphicx}
\usepackage{amsfonts}
\usepackage{amsmath}

\begin{document}

\title{TWO-PHOTON DECAY RATES OF TRUE NEUTRAL PSEUDOSCALAR
MESONS FROM DATA ON THEIR TRANSITION FORM FACTORS}

\date{\today}

\author{S.~Dubni\v cka}
\address{Institute of Physics, Slovak Academy of Sciences,
Bratislava, Slovak Republic}
\author{A.Z.Dubni\v ckov\'a}
\address{Dept. of Theoretical Physics, Comenius Univ., Bratislava,
Slovak Republic}
\author{A.~Liptaj}
\address{Institute of Physics, Slovak Academy of Sciences,
Bratislava, Slovak Republic}

\begin{abstract}
By means of the universal unitary and analytic model of
electromagnetic structure of hadrons the two-photon decay rates of
$P=\pi^0, \eta, \eta'$ mesons are determined in an alternative way
from data on their transition form factors.
\end{abstract}

\maketitle

\vspace{2cm}
\section {INTRODUCTION}

   There are three true neutral particles ($P=\pi^0, \eta, \eta'$) from the nonet of
pseudoscalar mesons, which decay into $2\gamma$. The  $\pi^0 \to
2\gamma$ channel with branching ratio $(98.798 \pm 0.032)\%$,
$\eta \to 2\gamma$ channel with branching ratio $(39.31 \pm
0.20)\%$ and $\eta' \to 2\gamma$ channel with branching ratio
$(2.10 \pm 0.12)\%$, corresponding to the following decay widths:
$\Gamma_{exp}(\pi^0 \to 2\gamma)=(7.84 \pm 0.56) eV$,
$\Gamma_{exp}(\eta \to 2\gamma)=(511.03 \pm 27.79) eV$ and
$\Gamma_{exp}(\eta' \to 2\gamma)=(4 305.00 \pm 424.95) eV$,
respectively.

    They are an average of repeated measurements at
different experiments exploiting the Primakoff effect and $e^+e^-
\to e^+e^-P$ process; in the case of $\eta'(958)$ also $e^+e^- \to
e^+e^-\pi^+\pi^-\gamma$, $e^+e^- \to e^+e^-\eta\pi^+\pi^-$ and
$e^+e^- \to e^+e^-\eta\pi^0\pi^0$ processes.

   In this contribution we demonstrate a more effective method of a
determination of $2\gamma$ decay widths of $\pi^0$, $\eta$ and
$\eta'$. More concretely, by a description of existing data on the
corresponding transition form factors (FFs) in space-like and
time-like regions simultaneously with the elaborate in
\cite{DDL04} Unitary and Analytic  ($U\&A$) model of $\pi^0$,
$\eta$ and $\eta'$ transition FFs.

    The main idea consists in the following: The behavior of the
meson-photon transition FF $F_{\gamma P}(Q^2)$ for $Q^2 \to 0$
\begin{equation}
lim_{Q^2 \to 0} F_{\gamma P}(Q^2) = \frac{1}{4\pi^2f_P}
\label{norm}
\end{equation}
can be determined from the axial anomaly in the chiral limit of
QCD, where $f_P$ is the meson weak decay constant and $Q^2 = -q^2
= -t$. However, in order to take into account the fact that
$f_\eta$ and $f_{\eta'}$ (unlike $f_{\pi}=93 MeV)$ are not directly
measurable quantities, employing the relation for the two-photon
partial width
\begin{equation}
\Gamma(P \to \gamma\gamma) = \frac{\alpha^2}{64 \pi^3 f_P^2} m^2_P
\label{decwidth}
\end{equation}
of the pseudoscalar meson $P$, one comes to a redefinition of the
FF norm
\begin{equation}
 F_{\gamma P}(0) = \frac{2}{\alpha m_P} \sqrt{\frac{\Gamma(P \to
 \gamma \gamma)}{\pi m_P}}
\label{redefnorm}
\end{equation}
to be expressed through the $\Gamma(P \to \gamma \gamma)$. So,
fitting all existing data on $F_{\gamma \pi^0}(t)$, $F_{\gamma
\eta}(t)$, $F_{\gamma \eta'}(t)$ in space-like and time-like
regions by the sophisticated $U\&A$ models simultaneously, one
finds normalization points values $F_{\gamma \pi^0}(0)$,
$F_{\gamma \eta}(0)$, $F_{\gamma \eta'}(0)$ and then, finally, by
means of (\ref{redefnorm}) the most reliable values of
$\Gamma(\pi^0 \to \gamma \gamma)$, $\Gamma(\eta \to \gamma
\gamma)$ and $\Gamma(\eta' \to \gamma \gamma)$.

\section {EXPERIMENTAL DATA ON PSEUDOSCALAR MESON TRANSITION FORM FACTORS}

    One of the first measurements of $\pi^0$, $\eta$ and $\eta'$
transition FFs in the space-like region was carried out by CELLO
Colab. \cite{Behrend91}, where really the $\pi^0$ transition FF in
the space-like region was observed for the first time.
An extension of the above-mention measured interval to higher
values of $Q^2$ was achieved by CLEO Collab. \cite{Gronberg98} to
be recently supplemented for $\pi^0$ up to $Q^2 = 34.36 GeV^2$ by
BABAR Collab. \cite{Aubert09}.
    For a measurements of $\pi^0$, $\eta$ and $\eta'$ transition FFs
in the time-like region commonly the annihilation processes
$e^+e^- \to \gamma P$ are used. Especially for $\pi^0$ and $\eta$
a lot of data was obtained on colliding $e^+ - e^-$ beams in
Novosibirsk by SND detector \cite{Achasov00}, \cite{Achasov03} and
by CMD-2 detector for $\eta$ transition FF \cite{Akhmetsin01} to
be corrected later on and published together with $\pi^0$
transition FF data in \cite{Akhmetsin05}.

\begin{figure}[htb]
\centering
\scalebox{0.4}{\includegraphics{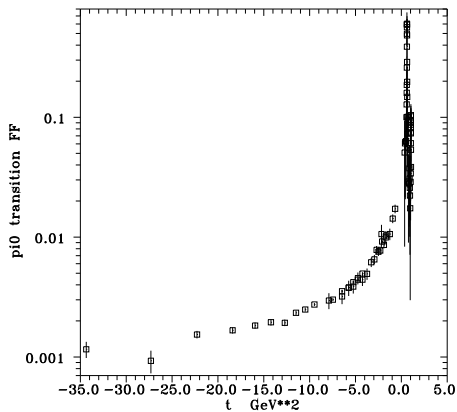}}
    \caption{\small{The data on $\gamma - \pi^0$ transition form factor.}}
    \label{fig:pi0trff}
\end{figure}

   Nevertheless, also here $1/3$ of the presented data on
$\sigma_{tot}(e^+e^- \to \eta \gamma)$ gives zero information on
$F_{\gamma \eta}(t)$ as there are only upper boundary estimations,
or the values to be charged by the error equal, even larger,
than the central value of the cross-section and this collection of
data have had to be carefully analyzed before our application.
   So, finally we are left with reliable (see Figs \ref{fig:pi0trff}-\ref{fig:etaprtrff}):
   \begin{itemize}
   \item
  81 exp. points on $\pi^0$ transition FF from the interval
   $-34.36 GeV^2 \leq t \leq 1.3535 GeV^2$;

  \item 58 exp. points on $\eta$ transition FF from the interval
   $-12.74 GeV^2 \leq t \leq 1.0685 GeV^2$;

   \item 56 exp. points on $\eta'$ transition FF from the interval
   $-15.3 GeV^2 \leq t \leq 0.48 GeV^2$,
   \end{itemize}

\begin{figure}[htb]
    \centering
        \scalebox{0.4}{\includegraphics{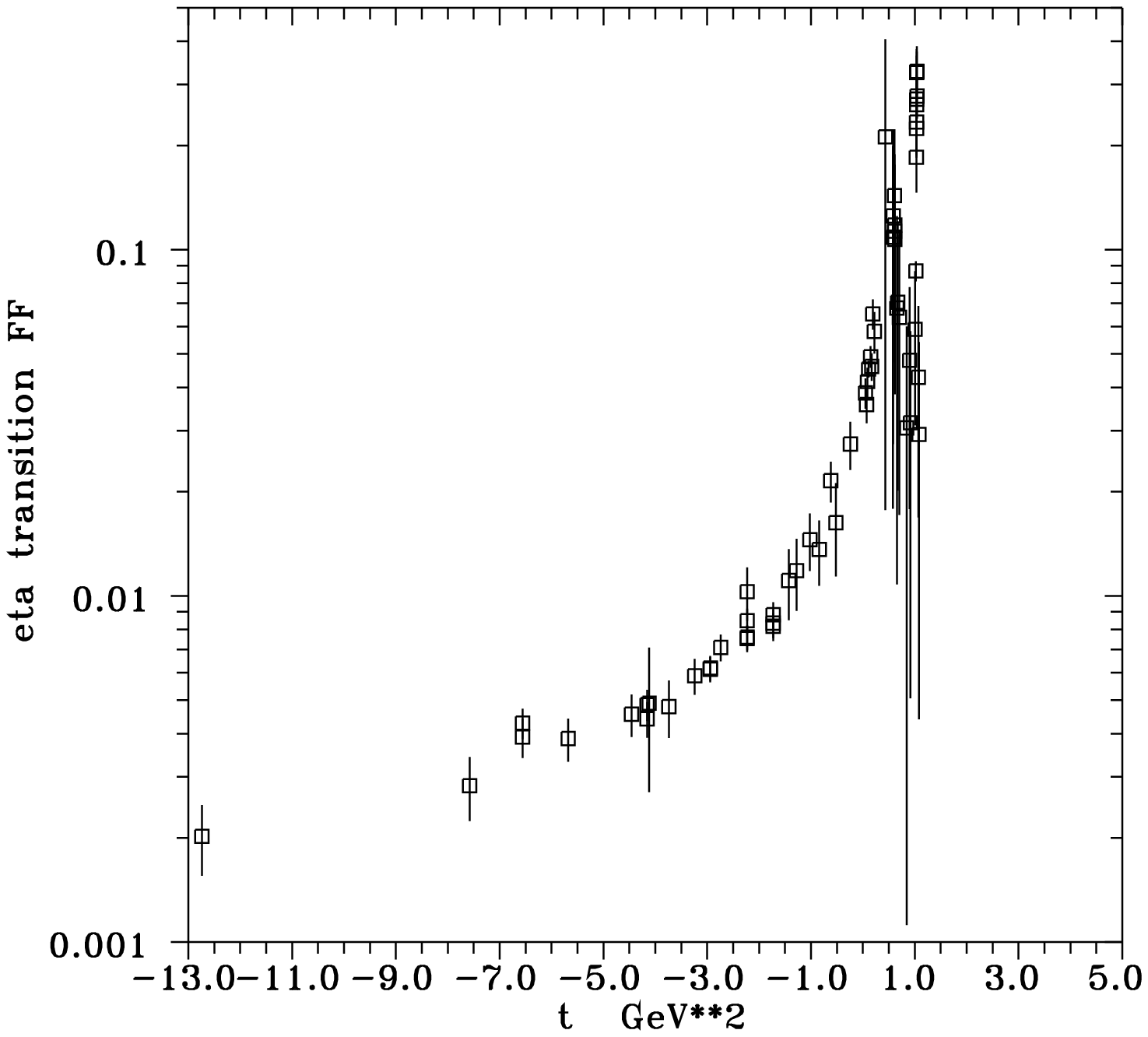}}
    \caption{\small{The data on $\gamma - \eta$ transition form factor.}}
    \label{fig:etaprtrff}
\end{figure}

\begin{figure}[htb]
    \centering
        \scalebox{0.4}{\includegraphics{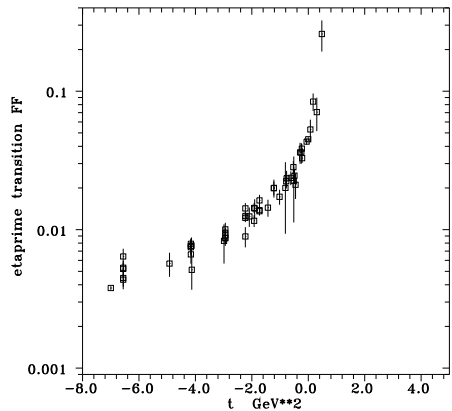}}
    \caption{\small{The data on $\gamma - \eta'$ transition form factor.}}
    \label{fig:etaprtrff}
\end{figure}
by means of which and the corresponding $U\&A$ models will be
$\Gamma(P \to \gamma \gamma)$ determined.

\section {UNITARY AND ANALYTIC MODEL OF PSEUDOSCALAR MESON TRANSITION FORM FACTORS}

   There is single FF for each $\gamma^* \to \gamma P$ transition
to be defined by a parametrization  of the matrix element of the
EM current $J_\mu^{EM} = 2/3\bar u \gamma_\mu u - 1/3 \bar d
\gamma_\mu d - 1/3 \bar s \gamma_\mu s$
\begin{equation}
<P(p)\gamma(k)\mid J_\mu^{EM} \mid 0> = \epsilon_{\mu \nu \alpha
\beta} p^\nu \epsilon^\alpha k^\beta F_{\gamma P}(q^2),
\end{equation}
where $\epsilon^\alpha$ is the polarization vector of $\gamma$,
$\epsilon_{\mu \nu \alpha \beta}$ appears as only the pseudoscalar
meson belongs to the abnormal spin-parity series.

    A straightforward calculation of $F_{\gamma P}(Q^2)$ behavior
in QCD is impossible and one can obtain \cite{BrodLep81} in the
framework of PQCD only the asymptotic behavior
\begin{equation}
\lim _{Q^2 \to \infty} Q^2 F_{\gamma P}(Q^2) = 2 f_P
\label{asympt}
\end{equation}
where $f_P$ is already explained the meson weak decay constant.

    Our intention is to achieve optimal description of all
$t<0$ and $t>0$ data on $F_{\gamma \pi^0}(t)$, $F_{\gamma
\eta}(t)$, $F_{\gamma \eta'}(t)$ always by one, however, distinct
for $\pi^0$, $\eta$, $\eta'$, analytic function explicitly known
on the real axis of $t$-plane from $-\infty$ to $+\infty$,
respecting all known properties of $F_{\gamma P}(t)$ like
\begin{itemize}
\item
the asymtotic behavior (\ref{asympt})
\item
the normalization (\ref{norm})
\item
the reality condition $F^*_{\gamma P}(t) = F_{\gamma P}(t^*)$
\item
analytic properties with the lowest branch point at
$t_0=m_{\pi^0}^2$
\item
unitarity condition, i.e. $Im F_{\gamma P}(t) \neq 0$ only from
the lowest branch point on the positive real axis of $t$-plane to
$+\infty$.
\end{itemize}
   The transition FF $F_{\gamma P}(t)$ is suitable to split into
two terms depending on the isotopic character of the photon
\begin{equation}
  F_{\gamma P}(t) = F^{I=0}_{\gamma P}(t) + F^{I=1}_{\gamma
  P}(t)
\end{equation}
where $F^{I=0}_{\gamma P}(t)$ can be saturated by only isoscalar
vector mesons and $F^{I=1}_{\gamma P}(t)$ can be saturated by only
isovector vector mesons, whereby both sets possess photon quantum
numbers.

   How much resonances will be considered?
   It is prescribed by the interval of data in $t>0$ region.
   The data on $\pi^0$ transition FF allow naturally to consider all
three ground state vector mesons $\rho(770)$, $\omega(782)$,
$\phi(1020)$ and also $\omega'(1420)$ and $\rho'(1450)$, in order
to obtain automatically normalized models for $F^{I=0}_{\gamma
P}(t)$ and $F^{I=1}_{\gamma P}(t)$.

   With the aim of obtaining comparable results for all three cases, the same number of
resonances is conserved also for $\eta$ and $\eta'$ transition FFs
and resonance parameters are fixed at the TABLE values.

   Then one can write down the following automatically normalized
five resonance VMD parametrization of transition FFs
\begin{eqnarray*}
 F^{I=0}_{\gamma P}(t)&=& \frac{1}{2}F_{\gamma
 P}(0)\frac{m_\omega'^2}{m_\omega'^2-t}\\
 &+&\{\frac{m_\omega^2}{m_\omega^2-t}
 -\frac{m_\omega'^2}{m_\omega'^2-t}\}(f_{\gamma
 P \omega}/f_\omega)\\
 &+&\{\frac{m_\phi^2}{m_\phi^2-t}
 -\frac{m_\omega'^2}{m_\omega'^2-t}\}(f_{\gamma
 P \phi}/f_\phi)
\end{eqnarray*}
\begin{eqnarray*}
 F^{I=1}_{\gamma P}(t)&=& \frac{1}{2}F_{\gamma
 P}(0)\frac{m_{\rho'}^2}{m_\rho'^2-t}\\
 &+&\{\frac{m_\rho^2}{m_\rho^2-t}
 -\frac{m_\rho'^2}{m_\rho'^2-t}\}(f_{\gamma
 P \rho}/f_\rho)
\end{eqnarray*}
suitable for a construction of $U\&A$ models of $F_{\gamma
\pi^0}(t)$, $F_{\gamma \eta}(t)$, $F_{\gamma \eta'}(t)$ transition
FFs.

    The analytic properties of $F_{\gamma P}(t)$ consist in
the assumption, that $F_{\gamma P}(t)$  is analytic in the whole
complex $t$-plane besides the cut on the positive real axis from
$t_0=m_{\pi^0}^2$ up to $+\infty$, as there is the intermediate
$\pi^0 \gamma$ state allowed in the unitarity condition of every
$\pi^0$, $\eta$ and $\eta'$ transition FF generating just the
lowest branch point $t_0=m_{\pi^0}^2$.
   On the other hand, just from the unitarity condition it
follows that there is an infinite number of higher branch points
on the positive real axis as there is allowed infinite number of
higher intermediate states in the unitarity condition of FFs under
consideration.
   In our model we restrict to two square-root cut
approximation of the latter picture, practically to be realized by
an application of the following nonlinear transformations
\begin{equation}
 t = t_0-\frac{4(t^s_{in}-t_0)}{[1/V-V]^2}\label{nonlin1}
\end{equation}
\begin{equation}
 t = t_0-\frac{4(t^v_{in}-t_0)}{[1/W-W]^2}\label{nonlin2}
\end{equation}
respectively, and subsequently also nonzero values of vector-meson
widths, $\Gamma_s \neq 0$ and $\Gamma_v \neq 0$, are established.

   The inelastic square-root branch points $t_{in}^s$ and
$t_{in}^v$ include in average contributions of all higher
important thresholds effectively and are left to be free
parameters of the constructed model.
   Variable $V$ ($W$) is conformal mapping
\begin{equation}
 V(t)=i\frac{\sqrt{q_{in}^s+q}-\sqrt{q_{in}^s-q}}{\sqrt{q_{in}^s+q}+\sqrt{q_{in}^s-q}}
\end{equation}
\begin{eqnarray*}
 q=[(t-t_0)/t_0]; \quad q_{in}^s=[(t_{in}^s-t_0)/t_0]
\end{eqnarray*}
of the four-sheeted Riemann surface in $t$-variable into one
$V$-plane ($W$-plane).
   As a result of application of the nonlinear transformations
(\ref{nonlin1}) and (\ref{nonlin2}) all VMD terms first give the
factorized forms
\begin{eqnarray*}
 \frac{m_i^2}{(m_i^2-t)}&=&(\frac{1-V^2}{1-V_N^2})
 \frac{(V_N-V_{i0})(V_N+V_{i0})(V_N-1/V_{i0})(V_N+1/V_{i0})}
 {(V-V_{i0})(V+V_{i0})(V-1/V_{i0})(V+1/V_{i0})}
\end{eqnarray*}
on the pure asymptotic term $(\frac{1-V^2}{1-V_N^2})$, independent
on the flavour of vector-mesons under consideration (however it
depends on the isospin) and carrying just the asymptotic behavior
of VMD model, and the so-called resonant term (the second one)
describing the resonant structure of VMD terms, which, however,
for $|t| \to \infty$ is going out on the real constant.
   The subindex $0$ means that still $\Gamma = 0$ of all vector
mesons is considered.
 \begin{figure}[htb]
    \centering
        \scalebox{0.4}{\includegraphics{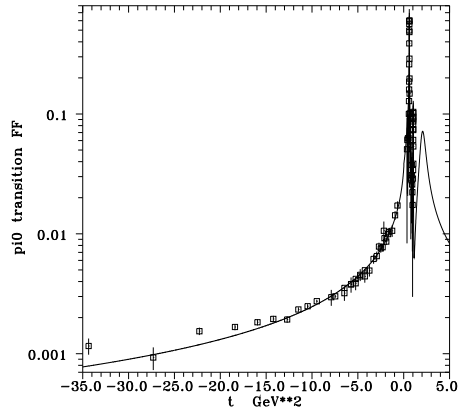}}
    \caption{\small{A description of data on $\gamma - \pi^0$ transition form factor.}}
    \label{fig:pi0th}
\end{figure}
    In order to demonstrate the reality condition $F^*_{\gamma
P}(t)=F_{\gamma P}(t^*)$ explicitly, one can utilize relations
between complex conjugate values of the corresponding zero-width
VMD model pole positions in $V$ ($W$) plane
\begin{equation}
 V_{\omega 0}=-V^*_{\omega 0}, \quad W_{\rho 0}=-W^*_{\rho 0}
\end{equation}
\begin{equation}
 V_{i0}=1/V^*_{i0}, \quad i=\phi, \omega'\quad W_{\rho'0}=1/W^*_{\rho'0}
\end{equation}
following from the experience that in a fitting procedure of
existing data on $F_{\gamma P}(t)$ such numerical value of
$t^s_{in}$,  $(t^v_{in})$ is found that
\begin{equation}
(m_i^2-\Gamma_i^2/4)<t^s_{in}, t^v_{in} \quad i=\omega, \rho
\end{equation}
\begin{equation}
 (m_j^2-\Gamma_j^2/4)>t^s_{in}, t^v_{in} \quad j=\phi, \omega',
 \rho'.
\end{equation}
   Finally, incorporating $\Gamma \neq 0$ by a substitution
\begin{equation}
 m_r^2 \to (m_r-i\Gamma_r/2)^2
\end{equation}
one comes to $U\&A$ model of $F_{\gamma P}(t)$ in the form
\begin{eqnarray*}
F^{I=0}_{\gamma
P}[V(t)]=(\frac{1-V^2}{1-V_N^2})^2\{\frac{1}{2}F_{\gamma
P}(0)H(\omega')\\
+[L(\omega)-H(\omega')]a_\omega\\
+[H(\phi)-H(\omega')]a_\phi \}
\end{eqnarray*}
\begin{eqnarray*}
F^{I=1}_{\gamma
P}[W(t)]=(\frac{1-W^2}{1-W_N^2})^2\{\frac{1}{2}F_{\gamma
P}(0)H(\rho')\\
+[L(\rho)-H(\rho')]a_\rho \}
\end{eqnarray*}
with
\begin{eqnarray*}
 L(\omega)=\frac{(V_N-V_\omega)(V_N-V^*_\omega)(V_N-1/V_\omega)(V_N-1/V^*_\omega)}
{(V-V_\omega)(V-V^*_\omega)(V-1/V_\omega)(V-1/V^*_\omega)}
\end{eqnarray*}
\begin{eqnarray*}
 H(i)=\frac{(V_N-V_i)(V_N-V^*_i)(V_N+V_i)(V_N+V^*_i)}
{(V-V_i)(V-V^*_i)(V+V_i)(V+V^*_i)}, i=\phi, \omega'
\end{eqnarray*}
\begin{eqnarray*}
 L(\rho)=\frac{(W_N-W_\rho)(W_N-W^*_\rho)(W_N-1/W_\rho)(W_N-1/W^*_\rho)}
{(W-W_\rho)(W-W^*_\rho)(W-1/W_\rho)(W-1/W^*_\rho)}
\end{eqnarray*}
\begin{eqnarray*}
 H({\rho'})=\frac{(W_N-W_{\rho'})(W_N-W^*_{\rho'})(W_N+W_{\rho'})(W_N+W^*_{\rho'})}
{(W-W_{\rho'})(W-W^*_{\rho'})(W+W_{\rho'})(W+W^*_{\rho'})}
\end{eqnarray*}
and normalization points $V(t)_{t=0}=V_N$, $W(t)_{t=0}=W_N$.
  It depends on the following five free parameters
\begin{eqnarray}
 t^s_{in}, t^v_{in}, a_j=(f_{\gamma P j}/f_j) \quad\quad j=\rho,
 \omega, \phi
\end{eqnarray}
which are found in an optimal description of existing data on
$\pi^0$, $\eta$ and $\eta'$ transition FFs, presented in previous
Figs. \ref{fig:pi0trff}-\ref{fig:etaprtrff}.
\begin{figure}[htb]
    \centering
        \scalebox{0.4}{\includegraphics{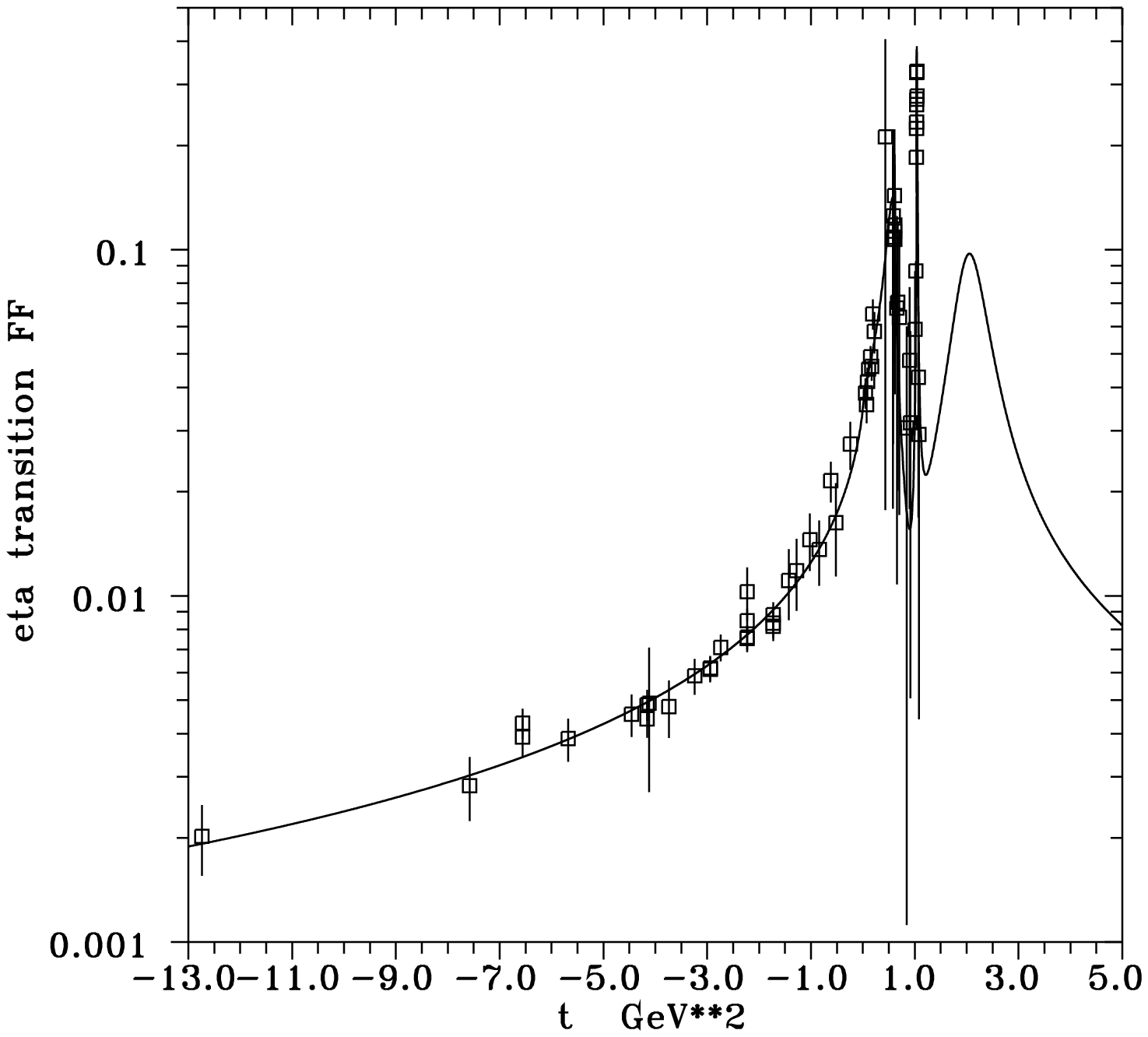}}
    \caption{\small{A description of data on $\gamma - \eta$ transition form factor.}}
    \label{fig:etath}
\end{figure}

\section{RESULTS}

    The constructed $U\&A$ model has been applied to optimal
 description of existing data and the results are as follows:
 for $\pi^0$: (see Fig.\ref{fig:pi0th})
\begin{eqnarray*}
 q^s_{in}&=&5.5210 \pm 0.0084; \quad q^v_{in}=5.6120 \pm 0.1414; \\
\quad a_\omega&=&0.0063 \pm 0.0013; \quad a_\phi=-0.0004  \pm 0.0001; \\
a_\rho&=&0.0212 \pm 0.0006; \quad F_{\gamma \pi^0}(0)=0.0352 \pm 0.0070; [m_\pi^{-1}]\quad \chi^2/ndf = 121/75 = 1.61,
\end{eqnarray*}

for $\eta$: (see Fig.\ref{fig:etath})
\begin{eqnarray*}
 q^s_{in}&=&6.7104 \pm 0.0190; \quad q^v_{in}=5.5006 \pm 0.0632;\\
 \quad a_\omega&=&0.0002 \pm 0.0014; \quad a_\phi=-0.0020 \pm 0.0003; \\
a_\rho&=&0.0250 \pm 0.0013; \quad F_{\gamma \eta}(0)=0.0348 \pm 0.0026; [m_\pi^{-1}] \quad \chi^2/ndf = 52/52 = 1.00
\end{eqnarray*}

 \begin{figure}[htb]
    \centering
        \scalebox{0.4}{\includegraphics{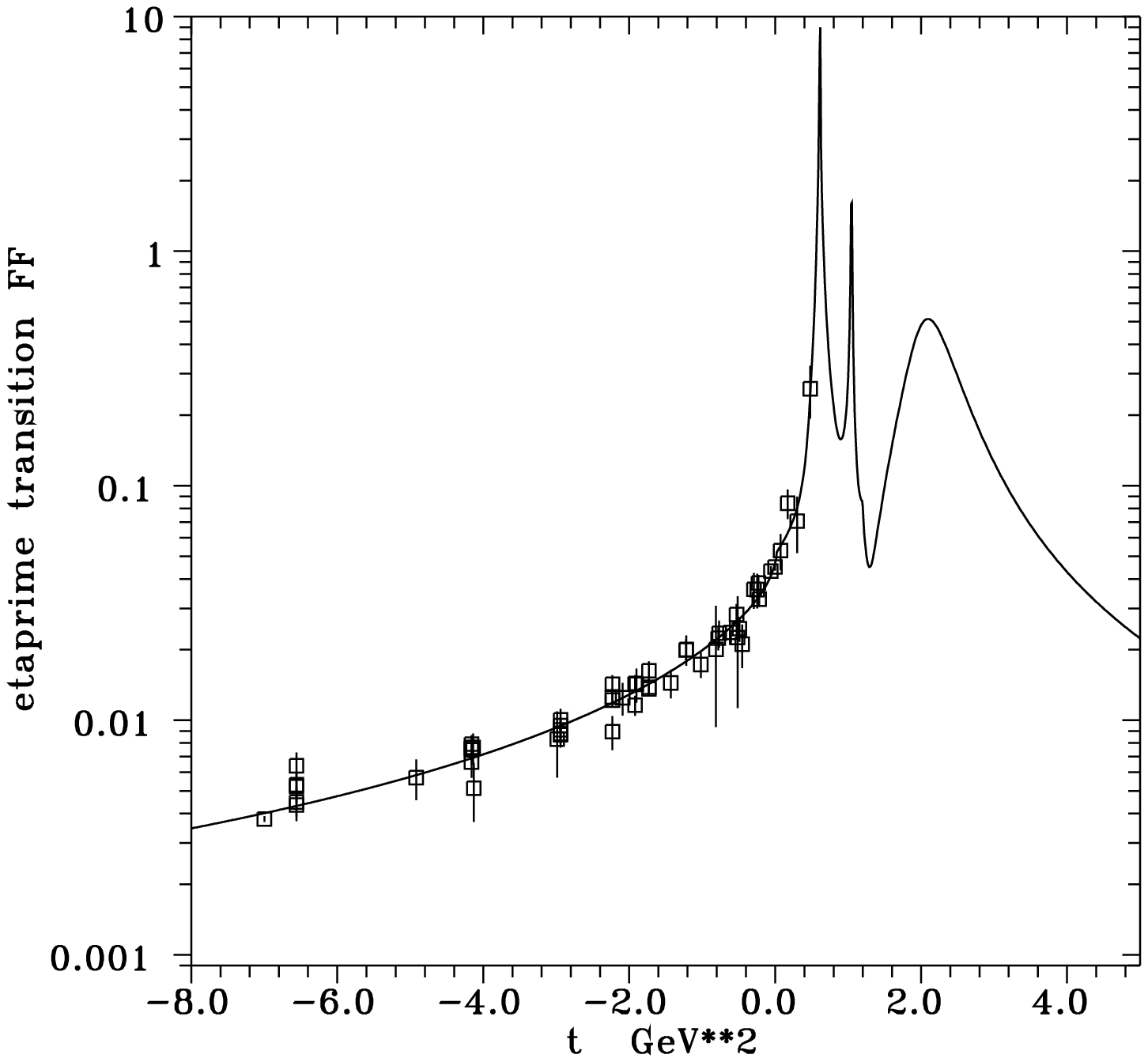}}
    \caption{\small{A description of data on $\gamma - \eta'$ transition form factor.}}
    \label{fig:etaprth}
\end{figure}

for $\eta'$: (see Fig.\ref{fig:etaprth})
\begin{eqnarray*}
 q^s_{in}&=&5.5366 \pm 0.0891; \quad q^v_{in}=7.7554 \pm 0.0158; \quad \\
a_\omega&=&-0.1134 \pm 0.0078; \quad a_\phi=0.0098 \pm 0.0091; \\
 a_\rho&=&0.1241 \pm 0.0026; \quad F_{\gamma \eta'}(0)=0.0469 \pm 0.0016 [m_\pi^{-1}];\quad \chi^2/ndf = 59/50 = 1.18
\end{eqnarray*}

    Finally, recalculated values of two-photon decay widths from the
obtained normalization points $F_{\gamma P}(0)$ are $\Gamma (\pi^0
\to \gamma \gamma) = (5.28 \pm 0.26) eV$, $\Gamma (\eta \to \gamma
\gamma) = (428.33 \pm 63.70) eV$ and $\Gamma (\eta' \to \gamma
\gamma) = (4142.88 \pm 274.01) eV$, to be compared with TABLE
values $\Gamma_{exp} (\pi^0 \to \gamma \gamma) = (7.84 \pm 0.56)
eV$, $\Gamma_{exp} (\eta \to \gamma \gamma) = (511.03 \pm 27.79)
eV$ and $\Gamma_{exp} (\eta' \to \gamma \gamma) = (4305.00 \pm
424.95) eV$, respectively.

\section{CONCLUSIONS}

   By an alternative method we have determined two-gamma
decay widths of $\pi^0$, $\eta$ and $\eta'$ pseudoscalar mesons.
The results are differing from the TABLE values, though our
results for $\pi^0$ and $\eta'$ are more precise. We are not
qualified to say, what results are more true. What can we say,
that if more precise data on $\pi^0$, $\eta$ and $\eta'$
transition FFs are measured, more precise values of pseudoscalar
mesons decay two-photon widths are obtained by our method.

The authors would like to thank the Slovak Grant Agency for
Sciences VEGA for support under the Grant No 2/0009/10.

\end{document}